\documentclass[preprint]{aastex}

\newcommand{\cmm}{cm$^{-2}$}

\newcommand{\kms}{km~s$^{-1}$}
\newcommand{\lya}{Ly$\alpha$}
\newcommand{\lyb}{Ly$\beta$}
\newcommand{\lyg}{Ly$\gamma$}

\newcommand{\zabs}{$z_{abs}$}

\newcommand{\hi}{{\rm H}~{\sc i}}

\newcommand{\di}{{\rm D}~{\sc i}}

\slugcomment{accepted for publication in the Astronomical Journal}
\shorttitle{\hi\ gas in higher density region of IGM}
\shortauthors{Misawa et al.}

\begin{document}

\title{H I gas in higher density regions of the intergalactic
 medium\altaffilmark{1}}

\footnotetext[1]{Data presented herein were obtained at the W.M. Keck
Observatory, which is operated as a scientific partnership among the
California Institute of Technology, the University of California and
the National Aeronautics and Space Administration. The Observatory was
made possible by the generous financial support of the W.M. Keck
Foundation.}

\author{Toru Misawa\altaffilmark{2}, 
        David Tytler\altaffilmark{3,4},
        Masanori Iye\altaffilmark{5,6},
        Pascal Paschos\altaffilmark{3},
        Michael Norman\altaffilmark{3},
        David Kirkman\altaffilmark{3,4},
        John O'Meara\altaffilmark{3,4},
        Nao Suzuki\altaffilmark{3,4}, and
        Nobunari Kashikawa\altaffilmark{5}
}

\altaffiltext{2}{Department of Astronomy and Astrophysics,
Pennsylvania State University, University Park, PA 16802;
misawa@astro.psu.edu}
\altaffiltext{3}{Center for Astrophysics and Space Sciences,
 University of California San Diego, MS 0424, La Jolla, CA
 92093-0424}
\altaffiltext{4}{Visiting Astronomer, W. M. Keck Observatory, which is
 a joint facility of the University of California, the California
 Institute of Technology, and NASA}
\altaffiltext{5}{National Astronomical Observatory, 2-21-1 Osawa,
 Mitaka, Tokyo 181-8588, Japan}
\altaffiltext{6}{Department of Astronomical Science, The Graduate
 University for Advanced Studies, 2-21-1 Osawa, Mitaka, Tokyo
 181-8588, Japan.}

\begin{abstract}
Using \hi\ absorption alone, we attempt to separate \hi\ absorption
lines in quasar spectra into two categories; HDLs (Higher Density
Lines) and LDLs (Lower Density Lines), and we discuss the difference
in their physical properties. We deblend and fit all \hi\ lines with
Voigt profiles, and make an unbiased sample of \hi\ lines covering a
wide column density range ($12 < \log N_{HI} < 19$ \cmm). To reduce
the influence of line blending, we simultaneously fit several Lyman
series lines. As a result of a two-point correlation analysis, we
found that higher column density \hi\ lines are clustering at $\Delta
v < 200$ \kms, while lower ones at $\Delta v < 100$ \kms. We define
HDLs as \hi\ lines with $15 < \log N_{HI} < 19$ \cmm\ and all \hi\
lines within $\pm$200 \kms\ of a line with $\log N_{HI} > 15$ \cmm,
and LDLs as others with $12 < \log N_{HI} < 15$ \cmm. We found that
the HDLs have smaller minimum $b$-values for a given column density
than the LDLs. This difference is successfully reproduced by our
Hydrodynamic simulation. The LDLs seem to be cool or shock-heated
diffuse IGM gas, while the HDLs are likely to be cooler dense gas near
to galaxies.
%The clustering trend of \hi\ lines is moderately consistent to the
%past work at larger velocity scales (e.g., $\Delta v$ $>$ 100
%\kms), while at smaller velocity scales (e.g., $\Delta v$ $<$
%50 \kms) \hi\ lines cannot be deblended completely and much higher 
%resolution spectra will be necessary.
\end{abstract}


\section{Introduction}
Quasars have been used as background sources, allowing the study of
objects that lie between us and them, most of which make Ly$\alpha$
absorption lines. To date, line-profile fitting analysis for \lya\
absorption lines has been carried out to study physical parameters
such as the velocity width and the column density (Pettini et
al. 1990; Carswell et al. 1991; Rauch et al. 1992; Hu et al. 1995 (H95
hereafter); Lu et al. 1996a (L96 hereafter); Kirkman \& Tytler 1997
(KT97 hereafter)).

Samples of \hi\ absorption lines usually contain (i) \hi\ lines
originating in the intergalactic diffuse gas, and (ii) \hi\
lines produced near or in intervening galaxies. The former could be
weak \hi\ lines, widely distributed as the \lya\ forest (hereafter we
call these ``LDLs'' or ``Lower Density Lines''), whereas the latter
are likely to be strong \hi\ lines clustered as metal absorption lines
(hereafter we call these ``HDLs'' or ``Higher Density Lines''). 
HDLs are thought to be caused by discrete clouds. On the other hand,
LDLs are probably arising from photoionized IGMs in the continuous density
fields that are broadened by Hubble flow (e.g., Rauch 1998; Kim et
al. 2002). Here,
the reader can think of density as column density $N_{HI}$, or volume
density, since numerical simulations show that the two are correlated
(Zhang et al. 1998). The HDLs are lines formed in or near to higher
density parts of the IGM. As the $\log N_{HI}$ increases, especially
above about 17, these regions are more likely to be in or near the
outer parts of galaxies, or clumps that will become galaxies. In
contrast, the LDLs arise in relatively lower density regions of the
IGM. 

There are studies which suggest that some \hi\ absorption lines have a
close relationship with galaxies near the line of sight, not only on
small scales at low-$z$ (Grogin \& Geller 1998; Penton, Stocke, \&
Shull 2002) but also on large scales at high-$z$ (Adelberger et
al. 2003). McDonald, Miralda-Escud\'e, \& Cen (2002) presented
theoretical predictions of the correlation between the \lya\ forest
transmitted flux and the mass of absorbers within $\sim$~5$h^{-1}$ Mpc
(comoving) of the line of sight.
% Adelberger et al. (2003) show that the intergalactic
% medium contains less neutral hydrogen within 0.5$h^{-1}$ Mpc of
% Lyman break galaxies. This trend can be reproduced in a
% hydrodynamical simulation only if the simulation includes a kinetic
% model of winds that can evacuate cavities around galaxies; the Lyman
% continuum radiation from galaxies does not work effectively (Croft
% et al. 2002).
Other spectroscopic and imaging observations have been carried out in
order to study the statistical relationship between the \lya\
absorbers and galaxies at low-$z$. The most important result is that
the \lya\ equivalent widths are anti-correlated with the projected
distance of the nearest galaxies within 500$h^{-1}$ kpc of the
galaxies (Lanzetta et al. 1995; Tripp, Lu, \& Savage 1998; Dav\'e et
al. 1999; Chen et al. 2001). Thus, some \hi\ absorbers seems to be
closely related with galaxies.

Thus, the analysis of HDLs and LDLs is indispensable for the detailed
investigation of physical properties of \hi\ gas. However, one of the
most serious problems in such an analysis is that the \lya\ absorption
lines are so heavily blended with each other that it is difficult to
separate and fit them individually. This problem is often seen in
strong \hi\ lines with $\log N_{HI} > 15$ \cmm.

In some past studies, the column densities of strong \hi\ lines have
been evaluated using the Lyman limit optical depth for LLSs, or the
total rest-frame equivalent width for DLA systems without fitting them
by Voigt profiles (e.g. Lanzetta et al. 1991; Petitjean et
al. 1993). These methods, however, evaluate only the total column
densities and Doppler parameters of the heavily blended \hi\ lines.

In this study, we attempt to fit all \hi\ lines with Voigt profiles,
construct an unbiased sample of \hi\ lines over a wide column density
range ($12 < \log N_{HI} < 19$ \cmm), and investigate their physical
parameters, such as column density, Doppler parameter, and clustering
properties. In order to separate the heavily blended (strong) \hi\
lines and fit them individually with Voigt profiles, we use not only
the \lya\ line but also higher Lyman series lines, such as \lyb\ and
\lyg, to improve the fitting accuracy.

This method has previously been applied to only a several LLS and DLA
systems (e.g. Songaila et al. 1994; Tytler, Fan, \& Burles 1996;
Wampler, Baldwin, \& Carswell 1996; Carswell et al. 1996; Songaila,
Wampler, \& Cowie, 1997; Burles \& Tytler 1998a,1998b, Burles,
Kirkman, \& Tytler 1999; Kirkman et al. 2000; O'Meara et al. 2001;
Kirkman et al. 2003) and the \lya\ forest in two quasar spectra (Kim
et al. 2002). Here we apply this fitting method to 40 quasar spectra
acquired with KECK + HIRES (Vogt et al. 1994).

Finally, we attempt to separate \hi\ lines into HDLs and LDLs, and
apply our statistical analysis to these two classes separately. We
confirm that the column density distribution and the Doppler parameter
distribution of LDLs are similar to the past results (H95; L96;
KT97). We found the most remarkable difference between HDLs and LDLs
in the plot of column density vs. Doppler parameter, which suggests
that \hi\ absorbers are not produced by a single phase (or a single
population). This difference is also reproduced by our Hydrodynamic
simulation. The summary of the other results and the detailed
description of each absorption system are presented in Misawa et
al. (2004).

We give a brief description of the observation in \S\ 2. In \S\ 3, we
explain how to evaluate the line parameters. In \S\ 4, we investigate
the physical properties of HDLs and LDLs, and compare the
observational results with the Hydrodynamic simulation. We summarize
our results in \S\ 5.

\section{Observations}
The sample quasars were originally selected from a survey taken for
measurements of the D/H (deuterium to hydrogen) abundance ratio. The
typical ratio of D/H is so small, 3 $\times 10^{-5}$ (Kirkman et
al. 2003 and references therein), that we can only detect \di\ lines
corresponding to \hi\ lines with large column densities, $\log N_{HI}
\geq 16.5$ \cmm. Given this fact, we observed quasars in which either
DLA systems or LLSs were detected. Here we examine KECK+HIRES spectra
of 40 quasars. We used a 1$^{\prime\prime}$.14 slit which provided a
resolution of 8.0 \kms. 
% Since some quasars were observed more than once, we combined the
% spectra in order to increase the S/N ratio. 
The spectra were extracted by  the automated program, MAKEE  package,
written by Tom Barlow.

\section{Preparation of uniform sample}
Instead of detecting all the \hi\ lines in the spectra of the 40
quasars, this study includes only the \hi\ lines with $\log N_{HI} >$
15 \cmm, and all other \hi\ lines within $\pm$ 1000 \kms\ of such \hi\
lines. The reason is described below.

\subsection{Detection of H I systems}
We at first searched for \hi\ lines with $\log N_{HI} > 15$ \cmm, in
the following way:
(1) we found \hi\ lines already discovered in DLA systems or LLSs in
earlier work (Sargent, Steidel, \& Boksenberg 1989; Lanzetta 1991;
Tytler 1982; Burles 1997),
(2) we checked for \hi\ lines corresponding to previously discovered
metal absorption systems (P\'eroux et al. 2001; Storrie-Lombardi et
al. 1996; Petitjean, Rauch, \& Carswell 1994; Lu et al. 1993; Steidel
\& Sargent 1992; Lanzetta et al. 1991; Barthel, Tytler, \& Thomson
1990; Steidel 1990; Sargent et al. 1980; Sargent, Steidel, \& 
Boksenberg 1988), and 
(3) we finally attempted to detect \hi\ lines ourselves that satisfied
both of the following conditions --- (i) the \hi\ line has a column
density of $\log N_{HI} \geq 15$ \cmm, and (ii) at least one higher
Lyman series, e.g. \lyb\ and \lyg\ lines, is present along with \lya\
in the observed wavelength range, which improves the reliability of
line fitting. We do not consider metal lines.

Metal absorption lines (Sargent et al. 1980; Young, Sargent, \&
Boksenberg 1982; Sargent, Steidel, \& Boksenberg 1988; Petitjean \&
Bergeron 1994; Churchill \& Vogt 2001; Pichon et al. 2003) including
those in DLA systems (Lu et al. 1996b) cluster over $\Delta v < 400$
\kms. It is then sufficient to detect all lines within $\pm$ 1000
\kms\ of the \hi\ lines with $\log N_{HI} > 15$ \cmm\ for the purpose
of studying the physical properties of weak and strong \hi\
lines. Where more than one \hi\ line with $\log N_{HI} > 15$ \cmm\ was
detected within this velocity window, the position of the line with
the largest column density (hereafter the ``main \hi\ component'') was
considered to be the center of the \hi\ system. As an example, we show
the velocity distribution of \hi\ lines for the system at \zabs =
3.321 in the spectrum of Q0014+8118 in Figure 1 (see also Burles,
Kirkman, \& Tytler 1999). With this method, we detected 86 \hi\
systems at $2.1 < z < 4.0$ in the spectra of 31 quasars, out of a
total of 40 examined.

\subsection{Evaluation of Line Parameter}
For the fitting of each absorption line with a Voigt profile, we fit
all the accessible lines in the Lyman series, helping us deblend \hi\
lines and make an unbiased sample of \hi\ lines.

In the process of line detection, we removed very narrow lines with
Doppler parameter of $b < 4.81$ \kms\ or FWHM $\sim$ 8.0 \kms, which
is the resolution of our spectra. We decided to ignore all lines with
$b < 15$ \kms\ as described later.
% Most of lines with $b < 15$ \kms\ are identified as metal lines,
% because the corresponding temperature of such narrow \hi\ lines is
% very low (T $\sim$ $1.4 \times 10^{4}$ K). In fact, such narrow \hi\
% lines rarely have been detected (e.g. H95; L96; KT97). We decided to
% ignore all lines with $b < 15$.
If the Voigt fitting allowed different solutions during the fitting
trials, the result with the minimum number of lines was chosen. This
occurs mainly in the saturated region for which we have only small
number of Lyman series information. If we fit the line profiles using
smaller number of components than the actual number, the
$b$-values could be overestimated. On the other hand, the column
density cannot be easily overestimated, because only a small rise of
the column density significantly changes the line profiles (e.g.,
damping profile) especially at the saturated regions.

Once the fitting model was applied to the absorption lines in some
region of the spectra, we used $\chi^{2}$ minimization to find the
model parameters which best fitted the observed spectrum. The formal
errors on the fitted parameters for lines with $b$ $<$ 30 \kms\ are
$\sigma(\log N_{HI})=0.09$ \cmm, $\sigma(b)=2.1$ \kms\ and
$\sigma(z)=2.5\times10^{-5}$. Before beginning our statistical
analysis, we prepared an unbiased sample of \hi\ lines, by eliminating
inappropriate \hi\ systems that have disadvantages described below.

\begin{description}
\item[(1) poor fitting due to gaps in echelle formatted spectra:]
  Echelle-formatted spectra sometimes have data gaps between echelle
  orders. This occurs frequently in the redder part of the spectrum
  (e.g., $\lambda$ $>$ 5000 \AA\ in our data). Spectral gaps appear
  every $\sim$ 100 \AA, and the widths of the wider gaps are no less
  than 20 \AA\ depending on the wavelength. If the Lyman series line
  damaged by the spectral gaps is \lya, the fitting accuracy for the
  system is very low. Therefore we removed \hi\ systems whose \lya\
  lines are affected by spectral gaps wider than 100\kms.
\item[(2) poor fitting due to strong DLA wings:] DLA (and sub-DLA)
  systems with large \hi\ column densities of $\log N_{HI}$ $\geq$ 19
  \cmm\ are unsuitable for our study, not only because their
  absorption profiles are damped so strongly that almost all weak
  components are blanketed by the damping wings, but also because the
  spectra around DLA lines are not normalized correctly due to the
  strong absorption features.
\item[(3) close proximity in redshift to the background quasars:]
  The \lya\ forest disappears in the regions redward of the \lya\
  emission lines of the quasars. Therefore we remove \hi\ systems at a
  distance of $\leq$ 1000 \kms\ from the \lya\ emission lines, because
  the asymmetrical distribution of their \hi\ lines complicates the
  clustering analysis described in \S\ 4.1.
\item[(4) overlapping with other \hi\ systems:] \hi\ system pairs,
  whose velocity windows of $\pm$1000 \kms\ are overlap, are
  also excluded, as their distribution of \hi\ lines could affect each
  other and the line clustering analysis is contaminated.
\end{description}

Finally, we have an unbiased sample of 973 \hi\ lines in 61 \hi\
systems. However the sample is not homogeneous because the S/N ratio
varies. The S/N ratios of the spectra are at least S/N $\simeq$ 11 per
2.1 \kms\ pixel and the mean value is S/N $\simeq$ 47 for \lya\ lines.

\section{Physical properties of H I lines}
We used the sample prepared in \S\ 3, to study the \hi\ line
parameters. Our analysis is similar to that of previous studies
(e.g. H95; L96; KT97), but with three key differences: (i) earlier
studies used all \hi\ lines detected in the quasar spectra, whereas we
use only \hi\ lines within $\pm$ 1000 \kms\ of the main components
with $\log N_{HI} > 15$ \cmm, (ii) our sample contains a number of
strong \hi\ lines ($\log N_{HI} > 15$ \cmm) in addition to weak lines
($\log N_{HI} < 15$ \cmm), and (iii) our sample covers a wide redshift
range, $2.0 \leq z \leq 4.0$, compared with those of the past studies,
$\Delta z \sim\ 0.5$.

\subsection{Clustering Properties}
Our sample contains not only LDLs but also HDLs. Dav\'e et al. (1999)
noted in their hydrodynamic simulations that galaxies tend to lie near
the dense regions that are responsible for strong \hi\
lines. Therefore, we attempted to classify the lines using their
clustering along the line of sight by constructing the two-point
correlation function (Sargent et al. 1980),
\begin{equation}
\xi(v) = \frac{N(v)}{N_{exp}(v)}-1,
\label{eqn:4.4}
\end{equation}
where $N(v)$ is the number of observed pairs at a velocity separation
of $v$, and $N_{exp}(v)$ is the number of line pairs expected if they
are randomly placed along the line of sight. $N_{exp}(v)$ is
calculated by Monte Carlo simulations. Webb (1987) found significant
clustering, $\xi (v)$ = $0.32\pm0.08$ over the velocity range 50 $< v
<$ 150 \kms\ at $1.9 < z < 2.8$. H95 confirmed the same trend with
$\xi (v)$ = $0.17\pm0.045$, over the same velocity range at $\langle z
\rangle = 2.8$. With a large sample of \hi\ lines with $\log N_{HI} >
13.8$ \cmm\ at $1.7 < z < 4.0$, Cristiani et al. (1997) also found
that the correlation function at $v \sim 100$ \kms\ increases with
$\log N_{HI}$. On the other hand, Rauch et al. (1992), L96, and KT97
did not find any clustering for similar velocity ranges at $2.7 < z <
3.4$, $\langle z \rangle = 3.7$, and $\langle z \rangle = 2.7$,
respectively.

In Figure 2, we show the number of \hi\ lines of our sample at a given
distance from the main components. The \hi\ line with relatively large
column densities tend to cluster around the main components, and have
a symmetrical distribution, suggesting that they are related to each
other. In contrast, the number of \hi\ lines with smaller column
densities decrease near the center of \hi\ systems.

Therefore we calculated $\xi (v)$ for some subsamples, varying the
column density ranges. At first, a number of artificial lines matching
the number of observed lines were randomly inserted into the 61 \hi\
system windows. Since the observed \hi\ systems always have main
components at $v=0$ \kms, we also placed an artificial line at the
center of each system in order to produce unbiased simulated
data. This process was repeated 250 times to obtain an average value
for $N_{exp}(v)$.

As a result of the analysis using all \hi\ lines, we found that $\xi
(v)$ shows a slight number excess at $\Delta v < 200$ \kms. The values
of $\xi (v)$ for \hi\ lines with various column density ranges in bins
with a spacing of 50 \kms\ at $50 < v < 200$ \kms\ are summarized in
Table 1. Column (1) is the range of column density, columns (2), (3),
and (4) are the values of $\xi(v)$ and $1\sigma$ Poisson errors for
velocity separations of 50 -- 100, 100 -- 150, and 150 -- 200 \kms,
respectively, and column (5) is the velocity width where the lower
$1\sigma$ deviation of $\xi (v)$ first goes below $\xi (v)=0$ over $v
> 50$ \kms. Column (6) is the number of \hi\ lines in each
subsample. We see non-zero $\xi(v)$ at velocity separations of $v \sim
100$ \kms\ for weak \hi\ lines, and $v \sim 200$ \kms\ for strong \hi\
lines. The correlation degree at $v$ = 50 -- 100 \kms\ was found to
have a maximum for \hi\ lines with $15 < \log N_{HI} < 19$ \cmm. This
is one reason why we chose to use $\log N_{HI} = 15$ \cmm\ as part of
our definition of HDLs.

Within $\sim$0.5$h^{-1}$ comoving Mpc of Lyman Break Galaxies (LBGs), 
the IGM 
contains less neutral hydrogen gas compared with the cosmologically
averaged value at $z$ $\sim$ 3 (Adelberger et al. 2003). Some
cosmological hydrodynamic simulations (e.g., Croft et al. 2002;
Kollmeier et al. 2003) showed that this effect could be produced
not by UV flux from the LBGs but by the galactic winds. This decreased
absorption was detected in Adelberger et al. (2003) because the lines
of sight to the quasars and the LBGs are different. In our spectra,
strong \hi\ lines (maybe produced in the intervening galaxies) absorb
most of the quasar flux within 50 \kms\ of their centers, 
which prevents us from detecting \hi\ deficit in the vicinity of the
strong \hi\ lines. 
%While our HDLs arise in regions with higher than
%the mean density, they are unlikely to be near the much higher
%densities expected near to LBGs. Perhaps the star formation has not
%started (or the activity is very weak) in weaker HDL systems of our
%sample.

\subsection{HDLs and LDLs}
We classified HDLs and LDLs with the following procedure. At first, we
regarded all \hi\ lines with $\log N_{HI} > 15$ \cmm\ as HDLs, because
the line clustering is stronger as $\log N_{HI}$ increases as
Cristiani et al. (1997) found, and we found above. But we emphasize
that the value of $\log N_{HI}$ = 15 \cmm\ is not strict. These strong
\hi\ lines may also be accompanied by weak \hi\ lines that are
physically associated with them. In fact, metal absorption lines
sometimes have a core-halo structure; the strongest line is at the
center of the absorption system, while the weak lines exist almost
symmetrically on both sides of the strong one (e.g. Lu et al. 1996b;
Prochaska et al. 2001; Misawa et al. 2003). Therefore, we defined a
velocity distribution width for HDLs, $v_{HDL}$. Since \hi\ lines with
relatively large column densities of $\log N_{HI}$ = 13 -- 19, 14 --
19, 15 -- 19, and 16 -- 19 \cmm\ are correlated with each other within
the velocity separation of $v \sim 200$ \kms, we adopted $v_{HDL} =
\pm 200$ \kms, and regard all \hi\ lines with $12 < \log N_{HI} < 15$
\cmm, within $\pm$ 200 \kms\ of \hi\ lines with $\log N_{HI} > 15$
\cmm\ as HDLs. Hereafter we call the lines with $\log N_{HI} < 15$,
lying within 200 \kms\ of a $\log N_{HI} > 15$ line, {\it halo} HDLs
in distinction from the original HDLs. Finally all the 973 \hi\ lines
were separated into 306 HDLs (including {\it halo} HDLs) and 667 LDLs
based on these criteria.

%Among 306 HDLs, 200 are weak lines with $\log N_{HI}$ $<$ 15 \cmm,
%which means that each strong HDLs (hereafter {\it core} HDLs) is
%accompanied by two {\it halo} HDLs on the average. The {\it core} HDLs
%are something like frames of the \hi\ cloud distributions, while {\it
%  halo} HDLs could be distributing around the frames. We, however,
%have to keep in mind that some {\it halo} HDLs could be blanketed in
%{\it core} HDLs, and that some LDLs could be located within 200 \kms\
%of {\it core} HDLs.

\subsection{Column density -- Doppler parameter relation}
Using the subsamples of HDLs and LDLs, we investigated the correlation
between the Doppler parameters and column densities. Our results are
summarized in Figure 3. 

In this figure, we also show lines with $5 < b < 15$ \kms. We find 62
LDLs and 16 HDLs with $12 < \log N_{HI} < 14$ \cmm\ and $5 < b < 15$
\kms, or 8\% of the 973 LDLs + HDLs with $b > 15$ \kms. The proportion
of HDLs to LDLs with $b < 15$ \kms\ (16/62 = 25.8\%) is similar to
that for lines with $b > 15$ and the same column density range
(142/583 = 24.4\%). Since we did not identify metal lines, many of
them could be metals, and since our spectra have a range of S/N, many
could be erroneous. Past studies that did consider these two effects
found very few \hi\ lines with these parameters (KT97; H95;
L96). Hence, we ignore all lines with $b < 15$ \kms.

When we ignore \hi\ lines with $b$ $<$ 15 \kms, we found a positive
correlation between the minimum Doppler parameter, $b_{min}$, and
column density not only for LDLs approximately fitted by
\begin{equation}
 b_{min} = 4.0 \times \log \left[\frac{N_{HI}~({\rm cm}^{-2})}{10^{12.5}}\right]+14.0 \hspace{5mm} {\rm km}\ {\rm s}^{-1}
\end{equation}
at $12.5 < \log N_{HI} < 15.0$ \cmm\ as seen in the past papers
(e.g. KT97), but also for HDLs fitted by
\begin{equation}
 b_{min} = 1.3 \times \log \left[\frac{N_{HI}~({\rm cm}^{-2})}{10^{12.5}}\right]+10.5 \hspace{5mm} {\rm km}\ {\rm s}^{-1}
\end{equation}
at $12.5 < \log N_{HI} < 19.0$ \cmm.
% Thus, HDLs have relatively small $b$-values compared to LDLs with the
% same column densities. Although the Doppler parameter, $b =
% \sqrt{b_{T}^{2}+b_{tur}^{2}}$, is broadened by both thermal broadening
% $b_{T}$ and non-thermal broadening $b_{tur}$, the minimum $b$-value
% $b_{min}$ is mainly influenced by thermal broadening because the
% corresponding pure thermal temperature (T $\sim$ 1 --- 3 $\times$
% $10^{4}$ K) is already the minimum acceptable value. We also attempted
% to pick HDLs that could be detected without line profiles of higher
% Lyman series (i.e. they would have been detected in the earlier
% studies without our fitting method), and marked them with filled red
% circles in Figure 3. None of them have Doppler parameters smaller than
% the minimum value given by eqn. (2), which means that the difference of
% the distributional trend between HDLs and LDLs is only apparent when
% the clustering \hi\ lines are deblended and separated into HDLs and
% LDLs. 
At $\log N_{HI} > 15$ \cmm, the HDLs often have $b$ values below the
extrapolation of eqn. (2) for the LDLs. We, however, cannot
compare the $b_{min}$ values of LDLs with those of HDLs directly at
$\log N_{HI} > 15$ \cmm, because there are no LDLs at that
region. Nonetheless, if we look at the region of $\log N_{HI}$ = 14
$-$ 15 \cmm, a difference in the distribution of LDLs and HDLs is
clear. At $\log N_{HI}$ = 14 -- 15, only 2 \%\
(2 of 91) LDLs are below the eqn. (2), while  20 \%\ (12 of 59) HDLs
are located there. We do not think this difference is just a 
statistical accident. At $\log N_{HI} < 14$
\cmm, we do not see any remarkable difference between them, which
could mean that there is little difference between the physical
properties of LDLs and {\it halo} HDLs.

\subsection{Comparison with simulation}
For weak \hi\ lines, the correlation between column density and
minimum Doppler parameter has already been reproduced by CDM
simulations. Zhang, Anninos, \& Norman (1995) performed the
hierarchical three-dimensional numerical simulation, and found the
minimum $b$-value is increasing slightly with column density in a
linear fashion. In order to mimic observational analysis more closely,
Zhang et al. (1997) synthesized absorption spectra, and extracted,
deblend, and fit the absorption features in generated spectra. The
recovered data exhibit a $b_{min}$ dependence on the column density
well fitted by $b_{min} = 5.5 \times \log [N ({\rm cm}^{-2})
  /10^{12.5}] + 12.8$ \kms, which is consistent with the KT97's and
our results for LDLs (eqn. (2)) shown in Figure 3.

We have performed a three-dimensional hydrodynamical simulation of
\lya\ clouds in a CDM dominated universe, and compared it with our
observational results. We used the following input parameters;
$\Omega_{b}$=0.04, $\Omega_{m}$=0.30, $\Omega_{\Lambda}$=0.70,
$H_{0}$=70 km~s$^{-1}$~Mpc$^{-1}$, $\sigma_{8}$=0.73, and power
spectrum slope $n = 1$. We chose
the size of the computational box to be $L$= 0.7$h^{-1}$ Mpc with the
effective grid resolution of 256$^{3}$. We used the Haardt \& Madau
(1996) quasar spectrum with the photo-ionization rate, $\gamma_{HI}$ =
$5.6\times10^{-13}\times(1+z)^{0.43}\times\exp[-(z-2.3)^{2}/1.95]$
ionizations per \hi\ atom per second in the optical thin
limit. Self-shielding is not included in the simulation. Figure 4
shows the number of \hi\ lines at a given distance from the main
component that are detected in our Hydrodynamic simulation at $z$ =
2.9 $-$ 3.0 (similar figure to Figure 2). The distribution of \hi\
lines are examined for three column density ranges, $13.5 < \log
N_{HI} < 14.5$ \cmm, $14.5 < \log N_{HI} < 15.5$ \cmm, and $15.5 <
\log N_{HI} < 16.5$ \cmm. \hi\ lines with relatively large column
density, $\log N_{HI} > 14.5$, tend to cluster around the main
components. On the other hand, weaker \hi\ lines show the decrease in
their number near the main components. These trends are similar to
those seen in the observation results, although the components with
$14.5 < \log N_{HI} < 15.5$ \cmm\ seem to be more concentrated in
velocity in the simulated spectra.

In Figure 5, we show scatter plots between $\log N_{HI}$ and $b$ from
the simulation. The difference of the distributions between HDLs and
LDLs that we found in the observational result are successfully
reproduced; strong HDLs with $\log N_{HI}$ $>$ 15 \cmm\ have Doppler
parameters smaller than the extrapolation of $b_{min}$ for LDLs
(eqn. (2)).
% Many weak \hi\ lines with $\log N_{HI} < 15$ \cmm\ are also located below
% the relation for LDLs. Some of them with $b < 15$ \kms\ are regarded
% as not \hi\ lines but metal lines in the observed result following the
% past studies. Others with $b > 15$ \kms\ and $\log N_{HI}$ = 13 --- 15
% \cmm\ could be HDLs located near strong \hi\ absorbers. 

One of the differences between the observation and the simulation is
that there is a blank region at $\log N_{HI} > 15$ \cmm\ and $b > 30$
\kms\ only in the simulation. It may be because absorption lines at
the region in the observation are not completely resolved even with
our improved fitting method (especially in the case that only a few
orders of Lyman series are available), and have large $b$-values
compared to the values that we might see with very high S/N ratio. The
minimum $b$-values, however, may be evaluated correctly even in such
cases. On the other hand, the failure of the line deblending rarely happen in
the simulation, because we have not included noise in the synthetic
spectra, and because the resolution
of our simulation, $256^{3}$ $0.7h^{1}$ Mpc ($\Delta x$ = 2.7 kpc), is
high enough compared to the minimum resolution ($\Delta x$ = 37.5 kpc)
necessary to evaluate
appropriate $b$-values (Bryan et al. 1999).

Another difference is that few narrow HDLs, whose Doppler parameter is
smaller than the value given by eqn. (2), are detected at $\log N_{HI}
< 15$ \cmm\ in the simulation (hereafter {\it weak-narrow}
HDLs). Here, we would like to emphasize that the $b$-values of \hi\
lines near the strong \hi\ absorbers could be over-estimated in the
simulation since we did not include self-shielding and shadowing
effects that can be effective for the HDLs. We discuss this in the
next section.

\section{Summary and Discussion}
In this paper, we, for the first time, separated \hi\ absorption lines
into HDLs and LDLs, and we find differences between them using 40
quasar spectra. 
For weak \hi\ lines (i.e. LDLs), the correlation between column
density and the minimum Doppler parameter have been often investigated
in both of the observation and the simulation. It is thought that this
relation arises by the mechanism that higher column density clouds
appear to be associated with denser regions in which the gas is
adiabatically compressed and heated (e.g. Kim, Cristiani, \& D'Odorico
2001, 2002). Kim, Cristiani, \& D'Odorico (2001, 2002) also suggested
that for weak \hi\ lines the slope
of the $\log N_{HI}$ $-$ $b_{min}$ relation became flatter as $z$
decreased at $z > 3$, though there was no such significant trend at
lower redshift $z \sim 2$. We, however, do not find any redshift
evolution in our sample (Figure 6 and 7).

The $\log N_{HI}$ $-$ $b_{min}$ relation for HDLs is clearly shallower
than the extrapolated relations for weak \hi\ lines in KT97 and Zhang
et al. (1997), while the relation for LDLs is good agreed with the
past results. This means that HDL absorbers have relatively small
Doppler parameters compared with the LDL absorbers at the same column
density.

Dav\'e et al. (1999) classify the \hi\ gas clouds into three phases;
(i) a cool low-density phase, (ii) a shock-heated intermediate-density
phase, and (iii) a cold dense phase in galaxies. The first phase with
relatively low column densities of $\log N_{HI} \leq 15$ \cmm\ make
the positive correlation between $\log N_{HI}$ and $b_{min}$, while
the last phase with large column densities makes an anti-correlation.

When we think about \hi\ clouds in the cold dense phase, the effect of
radiative cooling must be taken into consideration. Such dense clouds
with $\log N_{HI} \geq 17$ \cmm\ (i.e. LLSs and DLA systems) can
shield themselves against the background UV flux, which preserves more
\hi\ (i.e. shielding effect). A similar effect may happen in the low
column density clouds if they are located near the high column density
clouds (i.e. shadowing effect). These effects are important at least
at $z \geq 6$ (Nakamoto, Umemura, \& Susa 2001), although they have
not yet evaluated quantitatively at lower redshift because of high
complexity after the formations of first stars and quasars. Zhang et
al. (1995) also pointed out the importance of the shielding effect,
though they did not include the effect in their simulation, and nor
did we here. If plenty of \hi\ gas is preserved, the cooling of \hi\
gas becomes very effective in regions in photoionization equilibrium
(\hi\ line cooling is the most effective cooling factor at $T \sim\
10^{4}$ K), which could lower the temperature, and decrease the value
of the minimum Doppler parameter. Both of these effects will tend to
produce {\it weak-narrow} HDLs in the $\log N_{HI}$ $-$ $b$ plane,
which makes the distribution of \hi\ lines in the simulation more
consistent with the observation for the HDLs.

\acknowledgments
We would like to acknowledge to M. Chiba, R. Nishi, I. Murakami, 
M. Nagashima, and K. Okoshi who gave us valuable comments on the
theoretical interpretations. This work was supported in part by NASA
grant NAG5-13113. We wish to thank the anonymous referee for the
report, which improved the clarity of this presentation.

\clearpage

% Figure Legends
\figcaption[Figure 1]{Observed and modeled velocity map for 10 Lyman
  series lines of the \hi\ at \zabs\ = 3.321 in the spectrum of
  Q0014+8118. The lower line in each panel is the 1$\sigma$
  error. Tick marks denote the positions of \hi\ absorption lines, and
  the large one at the center is the position of the main component.}

\figcaption[Figure 2]{Distribution of 973 \hi\ lines, including the
  main components, in 100 \kms\ wide bins from the main components.}

\figcaption[Figure 3]{Doppler parameter vs. column density. 
  Open and filled circles are LDLs and HDLs, respectively. LDLs and
  HDLs with $b$ $<$ 15 \kms\ are marked with small circles. Solid line
  is the relation between the column density $\log N_{HI}$ and the
  minimum Doppler parameter $b_{min}$ in KT97 ($12 < \log N_{HI} < 15$
  \cmm), and dotted line is the same relation extrapolated up to $\log
  N_{HI} = 18$ \cmm. Dashed line is the same relation for HDLs in our
  sample. The pure thermal Doppler parameters corresponding to the
  temperatures, T = $10^{4}$, $5\times10^{4}$, $10^{5}$, and
  $2\times10^{5}$ K are shown with the thin dot-dashed horizontal
  lines.}

\figcaption[Figure 4]{Same as Figure 2, but the result from the
  Hydrodynamic simulation, except that the column density ranges of
  three subsamples are slightly different.}

\figcaption[Figure 5]{Similar to Figure 3, but from the Hydrodynamic
  simulation. Crosses are weak \hi\ lines with $\log N_{HI} < 15$
  \cmm. Open square and triangle are completely deblended ($b < 40$
  \kms) \hi\ lines with the column densities of $15 < \log N_{HI} <
  17$ \cmm\ and $\log N_{HI} > 17$ \cmm, respectively. In contrast
  with Figure 3, We do not use the open squares for all lines within
  200 \kms\ of lines with $\log N_{HI} > 15$ \cmm. Solid-and-dotted
  line and dashed line denote the relation between $\log N_{HI}$ and
  $b_{min}$ for LDLs (eqn. (2)) and HDLs (eqn. (3)) from the
  observational results.}

\figcaption[Figure 6]{Column density vs. Doppler parameter relation
  for HDLs at $z > 2.9$ and $z < 2.9$. We chose $z$ = 2.9 when
  dividing \hi\ sample into two subsamples because it makes the
  numbers of \hi\ systems in them nearly identical.}

\figcaption[Figure 7]{Same as Figure 6, but the result for LDLs.}

\clearpage

%%% Table 1 %%%
\begin{deluxetable}{cccccc}
\tablewidth{0pt}
\tablecaption{Clustering properties of \hi\ lines}
\tablehead{
\colhead{(1)} & 
\colhead{(2)} & 
\colhead{(3)} & 
\colhead{(4)} &
\colhead{(5)} &
\colhead{(6)} \\
\colhead{$\log N_{HI}$} &
\colhead{$\xi( 50 - 100)$\tablenotemark{a}} &
\colhead{$\xi(100 - 150)$\tablenotemark{a}} &
\colhead{$\xi(150 - 200)$\tablenotemark{a}} &
\colhead{$v$\tablenotemark{b}} &
\colhead{$n$\tablenotemark{c}} \\
\colhead{(\cmm)} & 
\colhead{} & 
\colhead{} & 
\colhead{} & 
\colhead{(\kms)} &
\colhead{}}
\startdata
12 --- 19 & $0.17^{+0.06}_{-0.06}$ & $ 0.08^{+0.05}_{-0.05}$ & $ 0.11^{+0.06}_{-0.06}$ & 200 & 972 \\
\hline
12 --- 13 & $0.52^{+0.28}_{-0.24}$ & $ 0.16^{+0.25}_{-0.21}$ & $-0.10^{+0.22}_{-0.18}$ & 100 & 244 \\
12 --- 14 & $0.32^{+0.08}_{-0.08}$ & $-0.00^{+0.07}_{-0.07}$ & $ 0.04^{+0.07}_{-0.07}$ & 100 & 716 \\
12 --- 15 & $0.21^{+0.06}_{-0.06}$ & $ 0.01^{+0.06}_{-0.06}$ & $ 0.06^{+0.06}_{-0.06}$ & 100 & 866 \\
12 --- 16 & $0.20^{+0.06}_{-0.06}$ & $ 0.05^{+0.06}_{-0.06}$ & $ 0.10^{+0.06}_{-0.06}$ & 100 & 933 \\
\hline
13 --- 19 & $0.19^{+0.08}_{-0.08}$ & $ 0.07^{+0.07}_{-0.07}$ & $ 0.15^{+0.08}_{-0.08}$ & 200 & 728 \\
14 --- 19 & $1.22^{+0.30}_{-0.30}$ & $ 0.79^{+0.31}_{-0.27}$ & $ 0.40^{+0.28}_{-0.24}$ & 200 & 256 \\
15 --- 19 & $4.03^{+1.49}_{-1.17}$ & $ 0.78^{+1.06}_{-0.70}$ & $ 1.02^{+1.09}_{-0.74}$ & 200 & 106 \\
16 --- 19 & $1.60^{+5.99}_{-2.15}$ & $ 3.76^{+6.28}_{-3.08}$ & $
3.55^{+6.00}_{-2.94}$ & 200\tablenotemark{d} & 39 \\
\enddata
\tablenotetext{a}{correlation and the 1$\sigma$ Poisson error
for the velocity separations of 50 -- 100, 100 -- 150, and
150 -- 200 \kms.}
\tablenotetext{b}{velocity width at which the lower $1\sigma$
deviation of $\xi(v)$ first goes below $\xi(v)$=0 over $v > 50$
\kms.}
\tablenotetext{c}{number of \hi\ lines.}
\tablenotetext{d}{we prefer to list $v$ = 200 \kms, though the lower
$1\sigma$ deviation of $\xi(v)$ goes below $\xi(v)$=0 at 50 $\leq v <$
100 \kms; this could be due to the small sample.}
\end{deluxetable}

%%% Table 2 %%%
%\begin{deluxetable}{cccc}
%\tablewidth{0pt}
%\tablecaption{Redshift evolution of \hi\ line clustering}
%\tablehead{
%\colhead{(1)} & 
%\colhead{(2)} & 
%\colhead{(3)} \\
%\colhead{$\log N_{HI}$} &
%\colhead{Redshift} &
%\colhead{$\xi(50 - 200)$} \\
%\colhead{(\cmm)} & 
%\colhead{} & 
%\colhead{} 
%}
%\startdata
%12 -- 19 & 2.1 -- 4.0 & 0.12$^{+0.03}_{-0.03}$\\
%         & 2.1 -- 2.9 & 0.14$^{+0.05}_{-0.05}$\\
%         & 2.9 -- 4.0 & 0.11$^{+0.04}_{-0.04}$\\
%12 -- 15 & 2.1 -- 4.0 & 0.09$^{+0.04}_{-0.04}$\\
%         & 2.1 -- 2.9 & 0.10$^{+0.06}_{-0.06}$\\
%         & 2.9 -- 4.0 & 0.10$^{+0.04}_{-0.04}$\\
%15 -- 19 & 2.1 -- 4.0 & 1.98$^{+0.64}_{-0.53}$\\
%         & 2.1 -- 2.9 & 2.07$^{+0.97}_{-0.76}$\\
%         & 2.9 -- 4.0 & 1.90$^{+0.96}_{-0.74}$\\
%\enddata
%\end{deluxetable}

% Figure 1
\clearpage
\begin{figure}
  \plotone{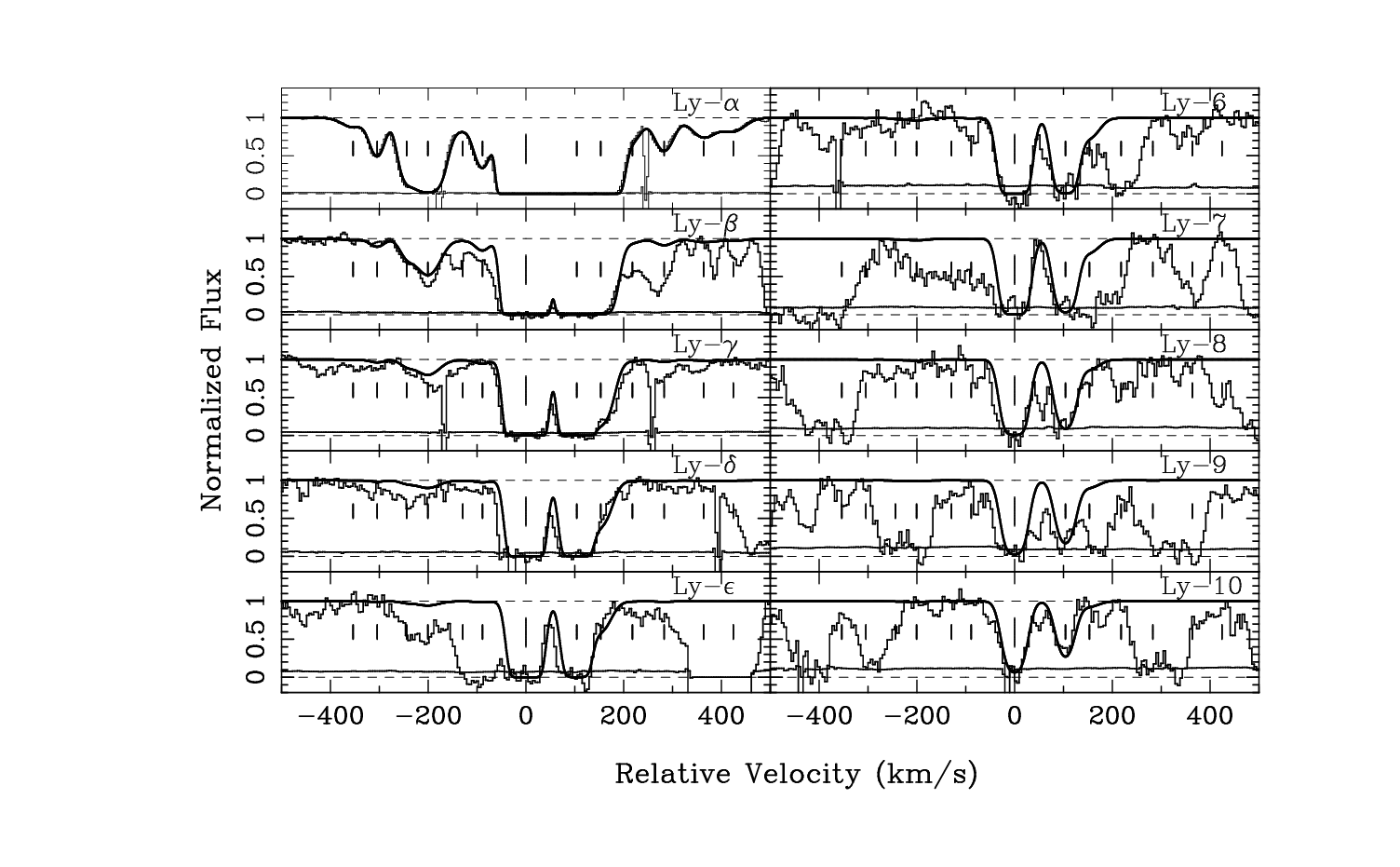}
\\Fig.~1~:~Velocity map of \hi\ system at \zabs\ = 3.321 in the
spectrum of Q0014+8118.
\end{figure}
\clearpage

% Figure 2
\clearpage
\begin{figure}
  \plotone{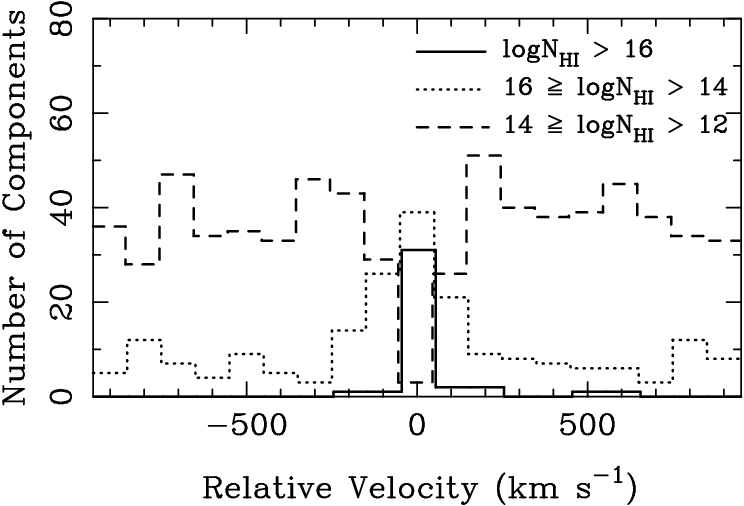}
\\Fig.~2~:~Distribution of \hi\ lines in 100 \kms\ bins from the
main components.
\end{figure}
\clearpage

% Figure 3
\clearpage
\begin{figure}
  \plotone{misawa.fig3.ps}
\\Fig.~3~:~Column density --- Doppler parameter relation for the
observation.
\end{figure}
\clearpage

% Figure 4
\clearpage
\begin{figure}
  \plotone{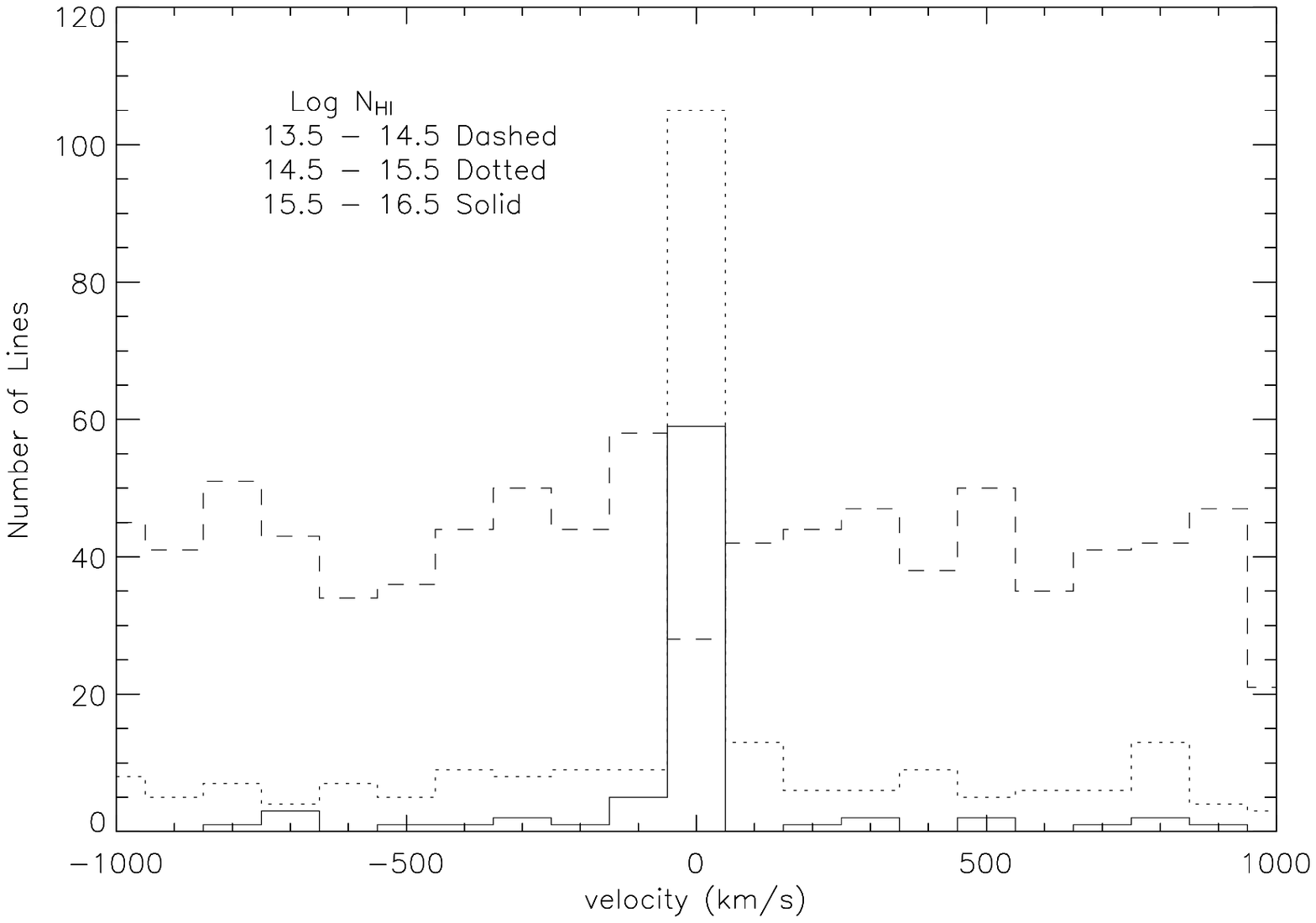}
\\Fig.~4~:~Same figure as Figure 2, but the result from the
simulation.
\end{figure}
\clearpage

% Figure 5
\clearpage
\begin{figure}
  \plotone{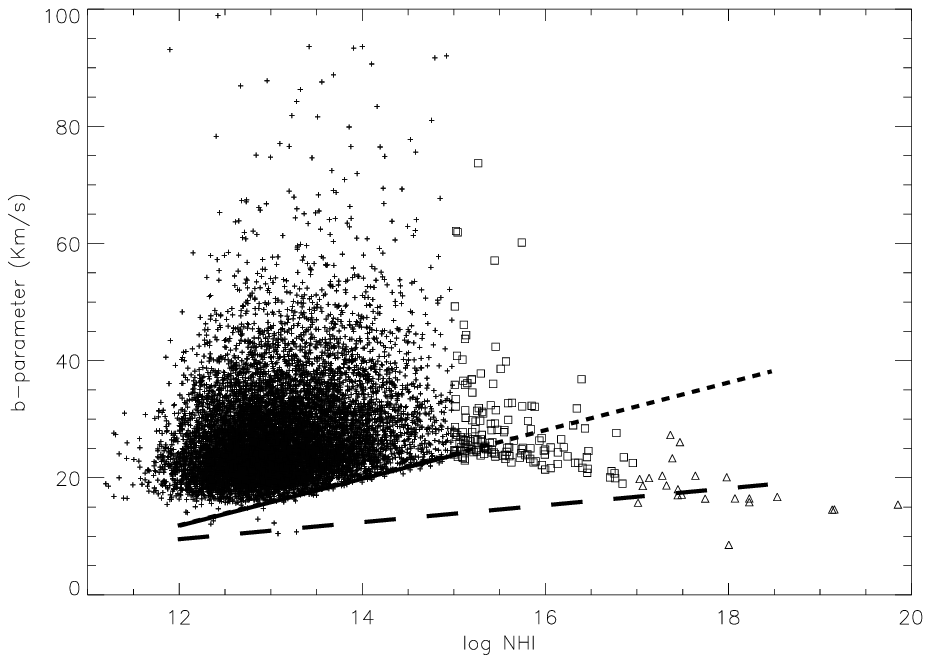}
\\Fig.~5~:~Same figure as Figure 3, but the result from the
simulation.
\end{figure}
\clearpage

% Figure 6
\clearpage
\begin{figure}
  \plotone{misawa.fig6.ps}
\\Fig.~6~:~Column density --- Doppler parameter relation for HDLs at
$z > 2.9$ and $z < 2.9$.
\end{figure}
\clearpage

% Figure 7
\clearpage
\begin{figure}
  \plotone{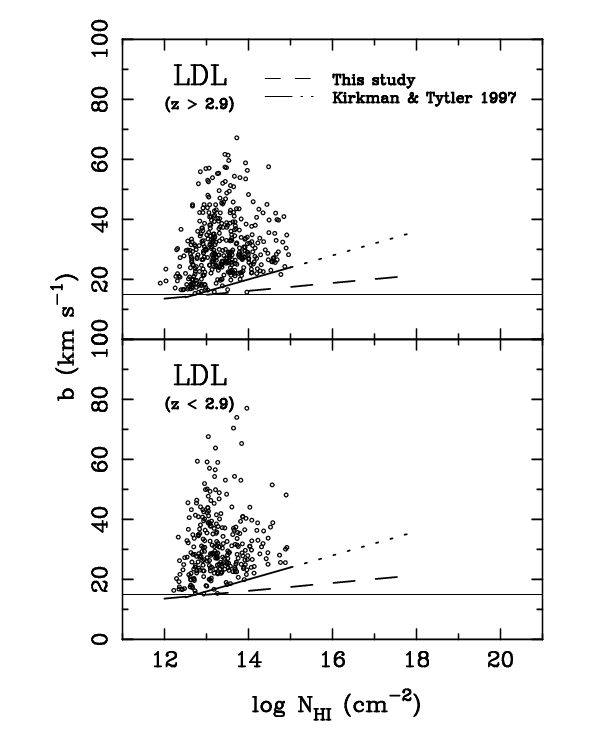}
\\Fig.~7~:~Same figure as Figure 3, but the result for LDLs.
\end{figure}
\clearpage

\end{document}